# Morphological Aspects of Porous Silicon Prepared by Photo-Electrochemical Etching

**Oday A. Abass, Sabah, M. Ali**

**School of Applied Sciences/University of Technology/ Department of Applied Physics ⁄ Branch of Laser ⁄ Baghdad-IRAQ**

## ABSTRACT

In this work the effects of coherent radiation (Laser) and incoherent radiation (Halogen lamp) during the electrochemical etching process on the structural characteristics of n-type PSi samples were investigated. The porosity values were measured by depending on the microstructure analyses and Gravimetric measurements. Surface morphology, layer thickness, pore diameter, pore shape, wall thickness and etching rate were studied by depending on Scanning electron-microscopic (SEM) images.

## 1. INTRODUCTION

Uhlir and Turner [1] first described the formation of porous silicon layers (PSLs) on silicon electrodes in hydrofluoric acid electrolytes under anodic bias in 1956.The discovery of a form of PSi with photoluminescence efficiency much higher than that of bulk Si has led to a flurry of research activity [2].Therefore this material has become popular between scientists and enters in various important applications. *Canham et al* [3], reported that the bulk Si can be made microporous (pore width ≤ 20 nm), mesoporous (pore width 20-500 nm) or macroporous (pore width > 500 nm), depending upon etching conditions. The photo-electrochemical etching takes advantage of the rectifying nature of the semiconductor/liquid junction; illumination increases the etching rate at n-Si [4]. *Thönissen et al* [5] illustrated that the illumination of the Si sample during or after the etching process known to be a further free etching parameter, which can be used to modify the P-Si layer morphology.

In this work, we report the influence of laser and halogen lamp illumination on the structural properties of n-type PSi samples during the etching process.

## 2. EXPERIMENT

Crystalline wafer of n-type Silicon with resistivity of 0.2 $\Omega$.cm, 508 µm thickness, and (111) orientation were used as starting substrates. The substrates were cut into rectangles with areas of    1 cm$^2$. The native oxide was cleaned in a mixture of HF and $H_2O$ (1:2). After chemical treatment, 0.1 µm-thick Al layers were deposited, by using an evaporation method, on the backsides of the wafer. Photo-electrochemical etching then performed in a mixture 49% (1:1) HF-Ethanol at room temperature by using a Pt electrode.

Current of 40 mA/cm$^2$ was applied for 10 min; Samples were illuminated in two ways in the first, using a halogen lamp (Philips Diachronic, 12 V, and 100 W) without filter and in the second by (810 nm, 2 W) diode laser. The etched area of sample was 0.6 cm$^2$ for both cases.

The structural properties: porosity, surface morphology, layer thickness, pore diameter, wall thickness, pore shape and etching rate measured. These measurements were achieved by depending on scanning electronic microscopy (SEM) images (Leo-1550). Voltage is given on each photo (5 kV). Samples tilted by 60 degrees, i.e. we are looking at an edge of the sample. Due to this tilt all layer thicknesses as measured from the photos have to be divides by sin(60°), i.e. multiplied by 1.155. The SEM magnification values were (5.45 k x–12.99 k x). These images were performed in Institute Fuer Biologic-and Nano-systems (IBN2), Germany.

## 3. RESULTS AND DISCUSSION

In this paper, we attempt achieving comparative study of morphological properties of PSi samples produced by photo-electrochemical etching using a diode laser and halogen lamp; therefore we have kept all parameters constant, except the source of illumination, as shown in tables (1). Figures (1,2) shows edge-view and top-view SEM images for formed PSi layer with PEC process at etching conditions as reported in table (1).

### 3.1. Etching rate

From SEM images, and using simple equation:

$$EtchingRat\,e = \frac{d}{t} \ldots\ldots\ldots\ldots\ldots\ldots\ldots\ldots\ldots\ldots(1)$$

Where $d$ is the layer thickness, $t$ is the etching time. We can calculate the etching rate values for formed PSi layer with laser beam and with halogen lamp radiation (white light) as shown in table (1).

We can note the following:

1. The etching rate of formed PSi layer with incoherent light was bigger than that with coherent light.

2. In both two cases the values of etching rate is various from position to other for same sample.

We attribute the first result to dependence of etching rate on generated electron-hole pairs and the latter depend upon the illumination power. In halogen lamp case, the charge carriers will generate in two modes, photo and thermal generation. For more details see *Ref.* [6,7].

The number of charge carriers (*n*) which was necessary to dissolve one silicon atom, was calculated: as

$$n = \frac{t m_{Si} J}{e d_{Si} V_D} \text{.....................................................(2)}$$

Where $V_D$ is the dissolved volume, e is the elementary charge; $m_{Si}$ is the mass of a Si atom, $d_{si}$ is the density of Si and *J* is the photocurrent density.

While in laser case, the photo-generated electron-hole pairs only, because the temperature rise is very little and seems too low to cause a sufficiently large thermal enhancement of the etch rate. For more details, see *Ref.* [8].

The number of charge carriers which was necessary to dissolve one silicon atom, was calculated as:

$$N_P = \frac{P}{hv} \text{.....................................(3)}$$

Where, P is the power of the laser, *hv* is the energy of photon, and *v* is the frequency of the laser and $N_p$ is the number of photons irradiated on silicon wafer per unit time.

The density of electron-hole pairs (G) generated per unit time given by:

$$G = \eta N_p \text{................................................(4)}$$

*η* is the quantum coefficient of Si, typically 0.78 at 810 nm.

By using equations (2 and 3) the required numbers of generated holes for etching process by halogen lamp radiation larger than that by laser beam. This result is reinforced by [6,7,8].

We can clarifying the second result especially for case of halogen lamp, that the etching rate were varying according to coefficient of absorption which is depend on the wavelength, and since the white light consist of range of wavelengths the photo-generated electron-hole pairs will create in various positions and in different depths.

While in laser case, there is one wavelength. The wafer Si with a smooth surface etched initially in three possible ways, two spreading out along the surface from the center, and one into the Si wafer.  The etching can start at various points in the center of the irradiated area because this area received most intense light; therefore, the resultant magnitude of etching rate in center is higher than that of other directions.  In fact, the etching rates will diminish continuously from the center of the laser beam to its periphery. For more details, see *Ref*. [9].

### 3.2. Surface morphology

The PSi samples formed with incoherent and coherent illumination have rough surfaces as shown in SEM images.  However, in same time, we can mark from these images; the surface of composed PSi layer with halogen lamp radiation is rougher than that formed in laser beam state. This phenomenon considered as result from the dependence of rough surface upon illumination power. This result supported by *Sejoon and Deuk* [6].

### 3.3. Porosity

The value of porosity ($P$) for both two cases calculated by the Gravimetric measurements as:

$$P = \frac{m_1 - m_2}{m_1 - m_3} \dots\dots\dots\dots\dots\dots\dots\dots\dots\dots(5)$$

Where, $m_1$ and $m_2$ are the weights of sample before and after etching process. $m_3$ is the weight of the sample after rremoving the PSi layer.

We observe from the reported values of formed PSi layer in table (1), the porosity value of fabricated PSi layer with halogen lamp radiation larger than that with laser beam. This result is support with surface roughness, which investigated from SEM images. We think the reason for this result is large number of pores in etched sample with halogen is possess large chance for the growing due to available bigger number of holes, while in case of laser this number is less. This result is supported by *Sejoon and Lehmann* [6,10].

### 3.4. Pore diameter and shape

The front side illumination, lead to a wide scatter in diameters and shapes of pore of the etched samples. For our samples, we can note from SEM images, the number of pores of the formed PSi layer with white light is decreases, while the diameter of the pores is increases. The reverse is true in laser beam state. The pore shape in first case is cylindrical while in the second state is rectangular. We attribute these results to dependence of distance between neighbored pores on the initial positions of the pores, which is relying on the local efficiency in capturing minority carriers, and the latter is depending on the penetration depth of incident radiation. The penetration depth relying on the wavelength of employed radiation. In additionally the carriers which are generated deeply in the bulk promote the pore growth at the tips, whereas near surface generation leads to lateral growth of the pores. Farther, in both cases we observe the pore diameter was varied over the length of the pore, this result refer to the important role of illumination intensity in determine the pore diameter. The degeneration of pore shape is most probably owing to minority carriers penetrating the region between the pores, which leads to enhanced etching of the pore walls. These results are consistent with [7, 11].

### 3.5. Layer thickness

From SEM images, we can note the following:

1- Incoherent light-assistance for PSi formation is producing layer thicker than that with coherent light.

2- In both two cases, the layer consists of two layers nanoporous and macroprous layer.

For constant parameters, we can explain the first result that the layer thickness found to be a linear function of wavelength and power of used illumination.While we believe that the macroporous layer obtained in the n-type Si wafer anodized under illumination is due to the fact that after a nanoporous layer of few micrometer is formed, light is not completely received by deepest layers and therefore the macroporous layer is formed in darkness. These results are reinforced by [7,11].

### 3.6. Wall thickness

From SEM images, the thickness of pore wall of the formed PSi layer with halogen lamp radiation is thinner than the thickness of pore wall of fabricated sample with laser beam. We attributed this result to dependence of thickness of pore wall on temperature. In case of halogen lamp radiation, the lost power as heat is 60% (for more details see *Ref.* [12]) from total power density (166.66 W/cm$^2$). and hence the temperature of sample will rising during the formation into a sufficient magnitude to cause a noticeable thermal enhancement for etching rate for walls between two pores, while we expect in laser case with power density (3.33 W?cm2) the temperature is not enough to cause any thermal enhancement in etching rate.. In addition, we can estimate according to wall thickness values that the walls in first case contain nano- particles with size smaller than that in second case. These results are sponsored by [11, 13,14].

**Table (1), Illustrated the structural properties of formed porous silicon layer at constant parameters. Current density (40 mA/cm², Etching time (10 min) and HF Consternation (1:1HF-Ethanol) at Room temperature.**

| Sample No. | Prosity (%) | Surface morphology | Layer thickness (µm) | Pore diameter (µm) | Pore shape | Wall thickness (µm) | Etching rate (µm/min) | Illumination source |
|---|---|---|---|---|---|---|---|---|
| 1 | 66 | roughness | 8 | $0.36 - 1.1$ | cylindrical | $0.18 - 2.18$ | 0.8 | Laser |
| 2 | 85 | more roughness | 24 | 0.57 - 2 | rectangular | $0.01 - 0.66$ | 2.4 | Halogen lamp |

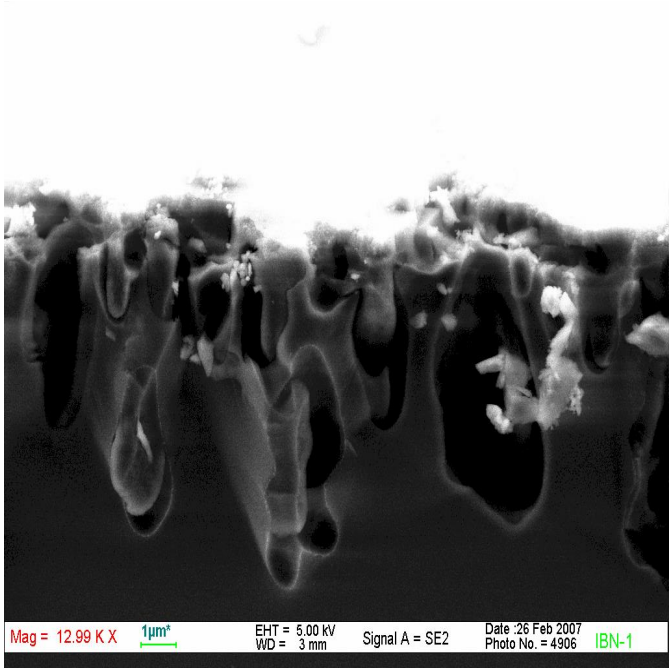
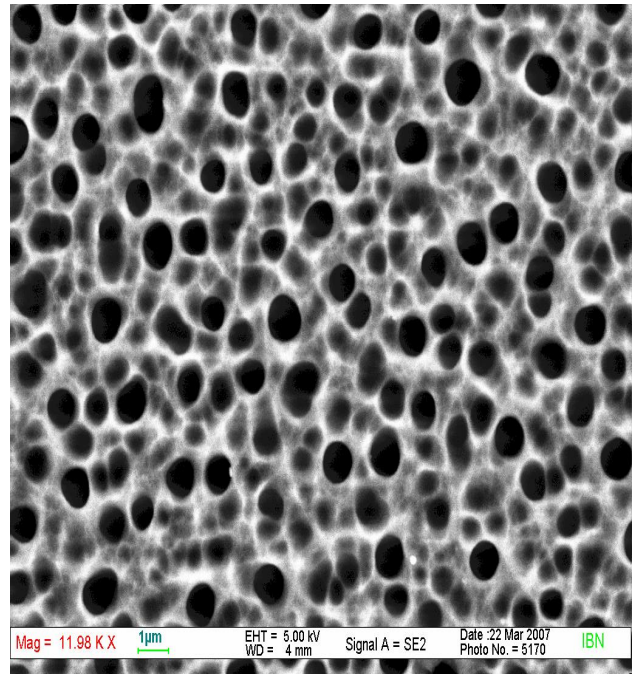

(a)                                                    (b)

**Fig. (1); SEM Image of formed porous silicon layer with Laser radiation at 10 min, 40mA/cm², 1:1HF-Ethanol. a- edge-view, b- top-view.**

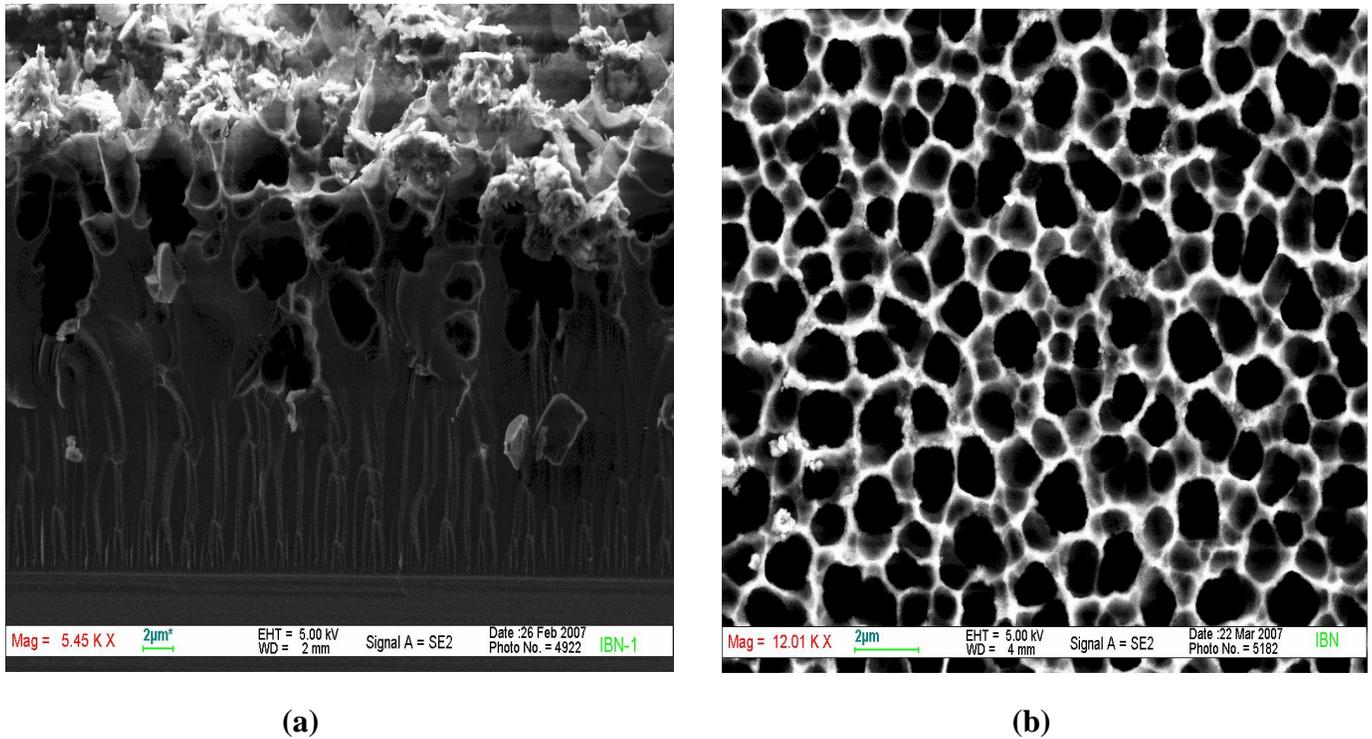

|  (a)  |  (b)  |

**Fig.(2); SEM Images of formed porous silicon layer with Halogen lamp radiation (white light) at 10 min, 40mA/cm$^2$, 1:1HF-Ethanol. a- edge-view, b- top-view.**

## 4. CONCLUSION

The obtained results show that the structural properties of PSi layer depends upon the wave length and power of employed illumination. In halogen lamp case, the surface roughness, layer thickness, porosity, and pore diameter are higher than these measured in the case of laser illuminated. The silicon nano-size in the case of incoherent radiation is smaller than that obtained in other case.

## ACKNOWLEDGEMENT

We are grateful to Dr. Hans Bohn for helpful discussions and useful references. This work supported by Institute Fuer Biologic-and Nano-systems (IBN2), Germany.